\documentclass[preprint,eqsecnum,aps]{revtex4}
\usepackage{graphicx}
\usepackage{amsmath}
\usepackage{latexsym}
\usepackage{float}
\begin{document}

\begin {center} 
\vskip 2.5 cm
{\LARGE {\bf A KINETICS DRIVEN  COMMENSURATE - INCOMMENSURATE TRANSITION}} \\
 
\vskip 2.0cm
 
{\large {\bf Abhishek Chaudhuri, P. A. Sreeram and Surajit Sengupta}} \\

\vskip 1.0cm 
{\normalsize {\bf
Satyendra Nath Bose National Centre for Basic Sciences \\
Block-JD, Sector-III, Salt Lake \\
Calcutta - 700098 \\
}}
\vskip 1 cm
{\bf Abstract}
\end {center}

\noindent
The steady state structure of an interface in an Ising system on a
square lattice placed in a {\em non-uniform} external field, shows a
commensurate -incommensurate transition driven by the velocity
of the interface. The non-uniform field has a profile with
a fixed shape which is designed to stabilize a flat interface, and is
translated with velocity $v_{e}$. For small velocities the interface is stuck
to the profile and is rippled with a periodicity which may be either
commensurate or incommensurate with the lattice parameter of the square
lattice. For a general orientation of the profile, the local slope
of the interface locks in to one of infinitely many rational directions
producing a devil's staircase structure. These ``lock-in'' or commensurate
structures dissappear as $v_e$ increases through a kinetics driven commensurate
- incommensurate transition. For large $v_e$ the interface becomes detached
from the field profile and coarsens with Kardar-Parisi-Zang exponents. The
complete phase~-diagram and the multifractal spectrum corresponding to these
structures have been obtained numerically together with several analytic
results concerning the dynamics of the rippled phases. Our work has
technological implications in crystal growth and the production of surfaces
with various desired surface morphologies.
\eject


\section{Introduction}
Commensurate-in~-commensurate (C-I) transitions\cite{Chaikin} have been 
extensively 
studied over almost half a century following early experiments on 
noble gases adsorbed on a crystalline substrate\cite{Kr} eg. Kr on graphite. 
Depending on coverage and temperature, adsorbates may show high 
density periodic structures the reciprocal lattice vectors (RLVs) 
of which are either a rational (commensurate) or irrational 
(in~-commensurate) multiple  of a substrate RLV. By changing external 
parameters (eg. temperature) one may induce phase-transitions between 
these structures. Recently, the upsurge of interest in the fabrication
of nano-devices have meant a renewed interest in this field following a
large number of experimental observations on ``self-assembled'' domain patterns
(stripes or droplets) on epitaxially grown thin films for eg. Ag films on
Ru(0001) or Cu-Pb films on Cu(111)\cite{films} etc. The alloy films often show
composition modulations in the lateral direction forming patterned 
superlattices. These self-assembled surface patterns may have potential 
applications in the field of opto-electronics, hence the interest. In general, 
the whole area of surface structure modification has tremendous technological 
implications including, for example, the recording industry where magnetic 
properties are intimately connected\cite{magnet} to surface structure.

Almost universally, C-I transitions may be understood 
using some version of the simple Frenkel~-Kontorova\cite{FK}(FK) model, 
which models them as arising from a 
competition between the elastic energy associated with the distortion of 
the adsorbate lattice and substrate~-adsorbate interactions. A 
complicated phase diagram involving an infinity of phases corresponding to 
various possible commensuration ratios (rational fractions) is obtained as 
a function of the two energy scales. In-between two commensurate structures 
one obtains regions where the periodicity of the adsorbate lattice is 
in~-commensurate. All C-I transitions are equilibrium 
transitions in the sense that at any value of the relevant parameters, 
the structures observed optimize a free~-energy. Indeed, despite its 
importance, the dynamical aspect of C-I transitions is a relatively unexplored 
domain.

In this paper, on the other hand, we discuss a C-I transition entirely driven 
by kinetics. We show that $(1)$ a simple Ising interface in a square lattice, 
held in place by a non-uniform external magnetic field, can have a variety of 
commensurate ``phases'' characterized by the {\em local} slope expressed in 
terms of the unit vectors of the underlying lattice and $(2)$ it is possible to 
induce transitions in-between these phases by externally driving the 
interface with the help of the field. The independent variables are 
therefore the average slope of the interface and its velocity both of which,
as we show, can be externally controlled. Preliminary results from this 
work has been published elsewhere\cite{physica,prl}.  

The dynamics of a 2-d Ising interface between the ``up'' and ``down'' spin
phases at low temperature T, in a (square) lattice driven by 
uniform external fields is a rather well studied\cite{Barabasi,Maj1} 
subject. The interface moves with a constant velocity, $v_{\infty}$, 
which depends on the applied 
field, $h$ and the orientation $\theta$ measured with respect to the underlying 
lattice. The interface is rough and coarsens with KPZ\cite{Barabasi} exponents 
$\alpha = 1/2$ and $\beta= 1/3$ where $\alpha$ and $\beta$  are the 
{\em roughness}  and  {\em dynamical} exponents respectively. 
We explore the possibility of driving such an interface with a pre-determined 
velocity $v_f$ using an external {\em non-uniform} field which changes sign 
following a sharp sigmoidal profile forcibly stabilizing a stationary, 
macroscopically flat interface at the region where the field crosses zero. 
We study systematically the structure and 
dynamics of this ``forced'' Ising interface as the field profile is moved without
change of shape at an externally controlable velocity $v_e$.
We show that for low driving velocities $v_e$, the interface velocity 
$v_f = v_e$ and the interface is stuck to the profile --- the ``stuck'' 
phase. For larger $v_e > v_{\infty}$, the interface detaches. In the stuck phase 
the interface, though macroscopically flat (i.e. $\alpha = \beta = 0$) is 
patterned on the scale of the lattice spacing. It is these patterns which 
we show, undergo a series of C-I transitions determined by the $v_e$ and the 
geometry characterized by the average slope of the interface in terms of 
the lattice vectors of the underlying square lattice.  

In the next section we introduce the model and briefly sketch the main results 
from a mean field treatment. In section III we 
map our interface dynamics to the dynamics of a one dimensional ``exclusion
process'' -- a system of hard core particles on a line. In section III-A and 
B we present our results for the ground state structure and the dynamics 
within this model. Finally, in section 
IV we conclude. 

\section{THE MODEL AND MEAN FIELD THEORY}

\begin{figure}[t]
\begin{center}
\includegraphics[width=6cm,height=9.2cm]{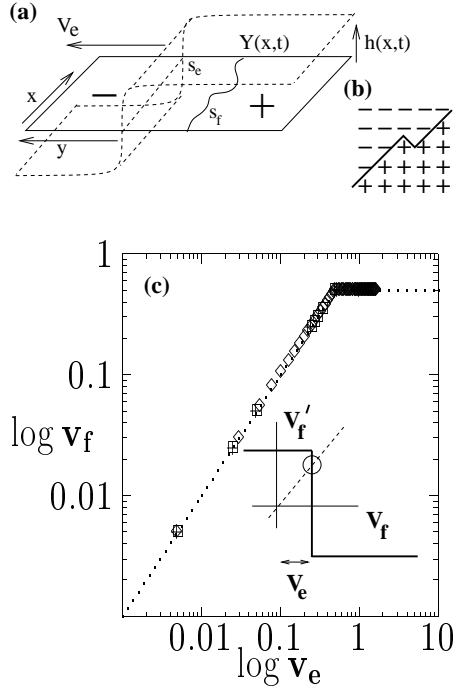}
\end{center}
\caption{
(a) An Ising interface $Y(x,t)$ (bold curved line) between regions of positive 
(marked $+$) and
negative (marked $-$) magnetization in an external, inhomogeneous field 
with a profile which is as shown(dashed line). The positions of the edge of the
field profile and that of the interface are labelled $S_e$ and $S_f$ respectively.
(b) A portion of the interface in a square lattice showing a corner. 
(c) The interface velocity $v_f$ as a function of the velocity of the
dragging edge $v_e$ for $N_s = 100(\Box), 1000(\Diamond), 10000(+)$
and $\rho = 0.5$. All the data ($\Box,\Diamond,+$) collapse on the
mean field solution (dashed line). Inset shows the graphical solution
(circled) of the self-consistency equation for $v_f$; dashed line
represents $v_f = v_f$.
}
\label{modvel}
\end{figure}

We consider here (see Fig.1) a one-dimensional interface $Y(x,t)$ between 
phases with 
magnetization, $\phi(x,y,t) > 0$ and $\phi(x,y,t) < 0$, in a 2-d square 
lattice obeying single-spin flip Glauber 
dynamics \cite{Men} in the limit $h/J , T/J \rightarrow 0$. 
Here $J$ is the Ising exchange coupling and $T$ the temperature.
An external non-uniform field   is applied
such that $h = {\it h_{max}}$ in the +ve  
and $-{\it h_{max}}$ in the -ve $\phi$ regions separated by
a sharp {\it edge}. The {\it edge} of the field (i.e. where the 
field changes sign) lies at $S_e$. The {\it front} or interface, $Y(x,t)$,  
separates up and down spin phases. The interface 
is the bold curved line (Fig.1) with the average position  $S_f$. 
When the edge is displaced with velocity ${\it v_e}$; 
the front,in response, travels with velocity ${\it v_f}$. Parts of the front 
which leads (lags) the edge of the 
field experience a backward (forward) force pulling it towards the edge.  
The driving force therefore varies in both space and time and depends on the 
relative position of the front compared to that of the edge of the dragging 
field. In the low temperature limit the 
interface moves solely by random corner flips\cite{Barabasi} 
(Fig.~\ref{modvel}(b)), the fluctuations  
necessary for nucleating islands of the minority phase in any region being
absent. We study the behaviour of the front velocity $v_f$ and the 
structure of the interface as a function of $v_e$ and orientation.

Naively, one would expect fluctuations of the interfacial coordinate $Y(x,t)$
to be completely suppressed in the presence of a field profile. This 
expectation, as depicted by our main results (Fig.\ref{modvel}, Fig.\ref{tanh}
and Fig.\ref{falpha}) is 
only partially true. While, as we show below, a mean field theory gives
the exact behaviour of the front velocity $v_f$ as a function of $v_e$
(Fig.\ref{modvel}); small interfacial fluctuations produce a dynamical phase 
diagram showing infinitely many dynamical phases (Fig \ref{tanh}) and 
dynamic phase transitions (Fig.~\ref{falpha}). 
For $v_e < v_{\infty}$ 
the interface is stuck to the profile $v_f = v_e$. The stuck phase
has a rich structure showing microscopic, ``lock-in'', commensurate
ripples. These dissappear at high velocities through a dynamical  
commensurate- incommensurate (C-I) transition.

Consider, first, a continuum coarse-grained description of the model
shown in Fig.~\ref{modvel} . We assume that for 
 $h/J , T/J \rightarrow 0$ the magnetization is uniform everywhere 
except near the interface, $Y(x,t)$ so that the magenetisation 
$\phi = \phi(y - Y(x,t))$. The field profile is given by
$h(y,t) =  h_{max} f(y,t)$ where $f(y,t) = \tanh ((y - v_{e}t)/\chi)$ 
and $\chi$ is the width of the profile (see Fig~\ref{modvel}(a)).
Using Model A dynamics\cite{Chaikin} for $\phi$ and integrating out all 
degrees of freedom except those corresponding to the interfacial position,
we obtain the effective dynamical equation satisfied by the interface,

\begin{equation}
\frac{\partial Y}{\partial t} = \lambda_{1} \frac{\partial^{2} Y}
{\partial x^{2}}-\lambda_{2} \Big(\frac{\partial Y}{\partial x} \Big)^{2} 
f(Y,t)-\lambda_{3} f(Y,t)+ \zeta(x,t)
\label{m-KPZ}
\end{equation}

\noindent
where $\lambda_{1}$,$\lambda_{2}$ and $\lambda_{3}$ are parameters. 
Note that Eq.\,(\ref{m-KPZ}) lacks Galilean invariance\cite{multif} 
$Y^{\prime} \rightarrow Y + \epsilon x,\>\> x^{\prime} \rightarrow x - 
\lambda_{2}\epsilon t, \>\> t^{\prime} \rightarrow t$. 
A mean field calculation amounts to taking $Y \equiv Y(t)$ i.e. neglecting 
spatial fluctuations of the interface and noise. 
For large times ($t \rightarrow \infty$), $Y \rightarrow v_{f}t$, where 
$v_{f}$ is obtained by solving the self-consistency equation;

\begin{eqnarray}
v_{f} &=& \lim_{t\to \infty} -\lambda_{3}\tanh \Big(\frac{(v_{f} - v_{e})t}{\chi} \Big) \nonumber \\
      &=& -\lambda_{3}\,{\rm sign}(v_{f} - v_{e})
\label{mft-v}
\end{eqnarray}

For small $v_e$ the only solution to Eq.\,(\ref{mft-v}) is 
$v_f = v_e$ and for  $v_{e} > v_{\infty}$, where 
$v_{\infty} = \lambda_{3}$ we get 
$v_{f} = \lambda_{3} = v_{\infty}$. We thus have a sharp transition\, 
(Fig.~\ref{modvel}(c)) from a region where the interface is stuck to the edge 
to one where it moves with a constant velocity.
How is this result altered by including spatial fluctuations of $Y$ ?
This question is best answered by mapping the interface problem to an 
assymmetric exclusion process\cite{Barabasi,AEP} and studying
the dynamics both analytically and numerically using computer simulations.

\section{ FLUCTUATIONS: THE EXCLUSION PROCESS}

The mapping to the exclusion process follows\cite{Maj1,AEP} by 
distributing  
$N_p$ particles among $N_s$ sites of a 1-d lattice. 
The particles are labelled $i = 1,2,.......,N_p$
sequentially at $t = 0$. Any configuration of the system is specified 
by the set of integers 
$\{y_i\}$ where $y_i$ denotes the location of the $i$th particle. In the
interface picture $i$ maps onto a  horizontal
coordinate ($x$ in Fig.~\ref{modvel}), and $y_i$ as the local height $Y(x)$. 
Each configuration $\{y_i\}$ defines a one-dimensional interface inclined
to the horizontal with mean slope $\tan \theta_f = 1/\rho$ where 
$\rho = N_{p}/N_{s}$. The $y_i$ satisfy the hard core constraint 
$y_{i+1} \geq y_i + 1$. The local slope near particle $i$ is given 
by $y_{i+1} - y_i$ and is equal to the inverse {\em local} density 
$\rho_i$ measured in a region around the $i^{\rm th}$ particle. 
Alternatively, one associates a vertical bond with 
a particle and a horizontal bond with a hole\cite{Maj1}, in which case, 
again, we 
obtain an interface with a slope $\tan \theta_f^\prime = \rho/(1-\rho)$. The
two mappings are distinct but equivalent. 
Periodic boundary conditions amount to setting
$y_{i+N_p} = y_i \pm N_s$.
Motion of the interface, by corner flips
corresponds to the hopping of particles. In each time step
($N_{p}$ attempted hops with particles chosen randomly and 
sequentially\cite{AEP}),
$y_i$ tends to increase (or decrease) by 1 with probability $p$ (or $q$); it
actually increses (or decreases) if and only if ${y_{i+1} - y_i > 1}$. 
The dynamics involving random sequential updates is known to indroduce the 
least amount of correlations among $y_i$ which enables one to 
derive exact analytic expressions for dynamical quantities using simple 
mean field arguments\cite{AEP}. 
The right and left jump probabilities $p$ and $q$ ($p+q = 1$) 
themselves depend on the relative position of the interface 
$y_i$ and the edge of the field profile $i/\rho + v_e t$. Note that this 
relative position is defined in a moving reference frame
with velocity $v_f(t)$, the instantaneous average particle velocity defined
as the total number of particles moving right per time 
step. We use a bias 
$\Delta_i(t) = p-q = \Delta\,{\rm sign}(y_i- i/\rho - v_e t)$ 
with $\Delta = 1$ unless otherwise stated. 
In addition to the front velocity $v_f$, we also examine the behaviour of 
the average position,
\begin{eqnarray}
<y(t)> = N_p^{-1}\sum_{i=1,N_p}y_i(t)
\end{eqnarray}
\noindent
and the width of the interface: 
\begin{eqnarray}
\sigma^{2}(t) &=& N_p^{-1}\sum_{i=1,N_p}<(y_i(t) - <y_i(t)>)^2>
\end{eqnarray}
as a function of time and system size $N_s$. Here, $<y_i(t)> = i/\rho + v_et$.
Angular brackets denotes an average over the realizations of the 
random noise.  Note that the usual particle hole symmetry for an exclusion 
process\cite{Barabasi,AEP}
is violated since exchanging particles and holes changes the relative position
of the interface compared to the edge.

We perform numerical simulations of the above 
model for $N_s$ upto $10^4$ to obtain $v_f$ for the steady 
state interface as a function of $v_e$ as shown in Fig.~\ref{modvel}(c).  
A sharp dynamical transition from an initially stuck interface with 
$v_f = v_e$ to a free, detached interface with 
$v_f = v_\infty = \Delta (1-\rho)$ is clearly evident as predicted 
by mean field theory. The detached interface coarsens with KPZ 
exponents\cite{physica}.
Note that, even though the mean field solution for $v_f(v_e)$ neglects 
the fluctuations present in our simulation, it is exact. The detailed 
nature of the stuck phase 
($v_f = v_e$ and $\sigma$ bounded) is, on the other hand,
considerably more complicated than the mean field assumption $Y(x,t)=Y(t)$.
\begin{figure}[t]
\begin{center}
\includegraphics[width=8cm]{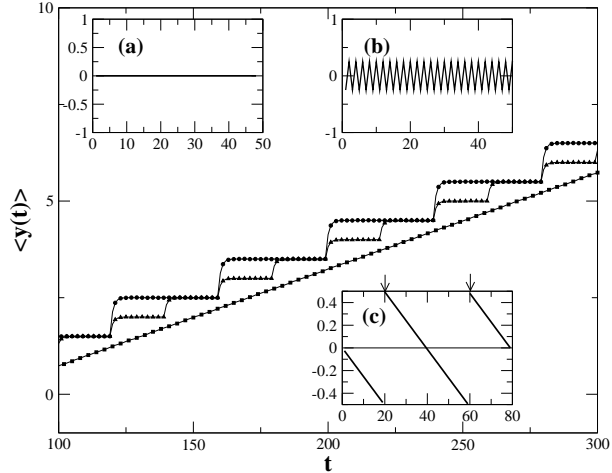}
\end{center}
\caption{ Variation of $<y(t)>$ with $t$ for 
$v_{e} = 0.025$ and $p = 1.0$. Lines denote analytic
results while points denote Monte Carlo data for $\rho = 1/5$ (uppermost curve),
$2/5$ and an incommensurate $\rho$ near $1/3$.
Inset (a)-(c) shows the corresponding ground state 
interfaces ($y_i-i/\rho$). The arrows in (c) mark the positions of two 
discommensurations.
}
\label{hops}
\end{figure}

\subsection{Ground state structure and the Devil's staircase}

The ground state of the interface in the presence of a stationary 
($v_e = 0$) field profile is obtained by minimizing 
$E = \sum_i (y_i- i/\rho - c)^2$ with respect to the set $\{y_i\}$ 
and $c$. This maybe shown from Eq.(\ref{m-KPZ})
by neglecting the terms non-linear in $\partial Y/\partial x$; the resulting 
equation of motion, for small deviations of $Y$ from the edge may be derived 
from the effective Hamiltonian $E$. 
The form of $E$ leads 
to an infinite range, non-local, repulsive, interaction between particles 
in addition to hard core repulsion and the minimization is subject to the 
constraint that  
$y_i$ be an integer. For our system, the result for the energy may be obtained
exactly for density $\rho = m/n$, an arbitrary rational fraction.
For even m we have the following expression.

\begin{eqnarray}
E &=& \frac{1}{m} \sum_{j=1,m/2} [(\frac{j}{m}+c)^2 + 
                    (\frac{j}{m}-c)^2 ] + \nonumber\\  
  & &                (\frac{1}{2}-c)^2 + c^2 
\nonumber\\
&=& \frac{1}{6}(\frac{1}{2}-\frac{1}{m})(1-\frac{1}{m}) +
 \frac{1}{4m} -\frac {1}{4m^2}
\end{eqnarray} 

Where in the last equation we have minimized the expression with respect to 
$c$.

\noindent
Similarly for odd m we have the following expression for energy,
\begin{eqnarray}
E &=& \frac{1}{m} \sum_{j=1,m/2-1} [(\frac{j}{m}+c)^2 + 
                    (\frac{j}{m}-c)^2 ] + c^2\nonumber\\  
\nonumber\\
&=& \frac{1}{12}(1 - \frac{1}{m^2})
\end{eqnarray} 

\begin{figure}[H]
\begin{center}
\includegraphics[width=10cm]{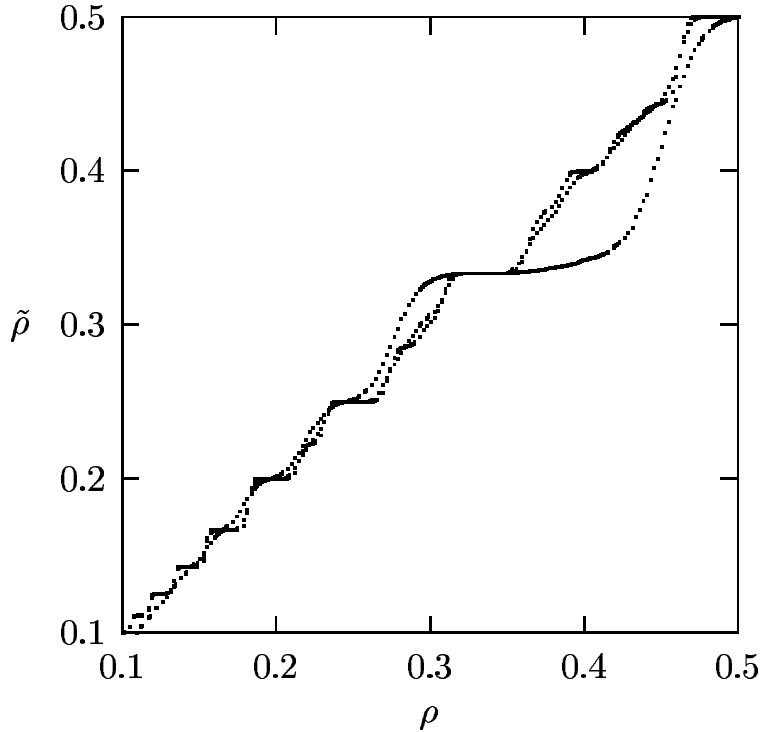}
\end{center}
\caption{
Devil's staircase structure in a plot of $\tilde\rho$ Vs $\rho$ for small 
velocities ($v_e = 0.05$) and $\chi = 0.01, 1, 5$.  
}
\label{tanh}
\end{figure}

\noindent
The resulting ground state profiles 
are shown in Fig. ~\ref{hops}(insets). The lower bound for $E(\rho)$ is zero 
which is the energy for all $\rho = 1/n$.
For irrational $\rho$ the energy is given by 
$\lim_{m\to \infty} E(m/n) = 1/12$ which constitutes an upper bound. 
For an arbitrary $0\,<\,\rho\,<\,1$ the system ($\{y_i\}$) therefore prefers 
to distort, conforming within local regions, to the nearest low-lying 
rational slope $1/\tilde{\rho}$ interspersed with ``discommensurations'' 
of density $\rho_d = |\rho - \tilde{\rho}|$ and sign +ve (-ve)
if these regions are shifted towards (away) from each other by $1$.  
A plot of $\tilde{\rho}(\rho)$ shows a ``Devil's staircase'' 
structure\cite{Chaikin}. In order to observe this in our simulation
we analyze the instantaneous distribution of the local density of the 
particle~-hole system to obtain weights for various simple rational fractions.  
A time average of these weights then give us the most probable density
$\tilde{\rho}$ --- distinct from the average $\rho$ which is constrained 
to be the inverse slope of the profile. Increasing the width $\chi$ of the 
external field profile away from zero gradually washes out this Devil's 
staircase structure (Fig.~\ref{tanh}).

\begin{figure}[H]
\begin{center}
\includegraphics[width=14cm]{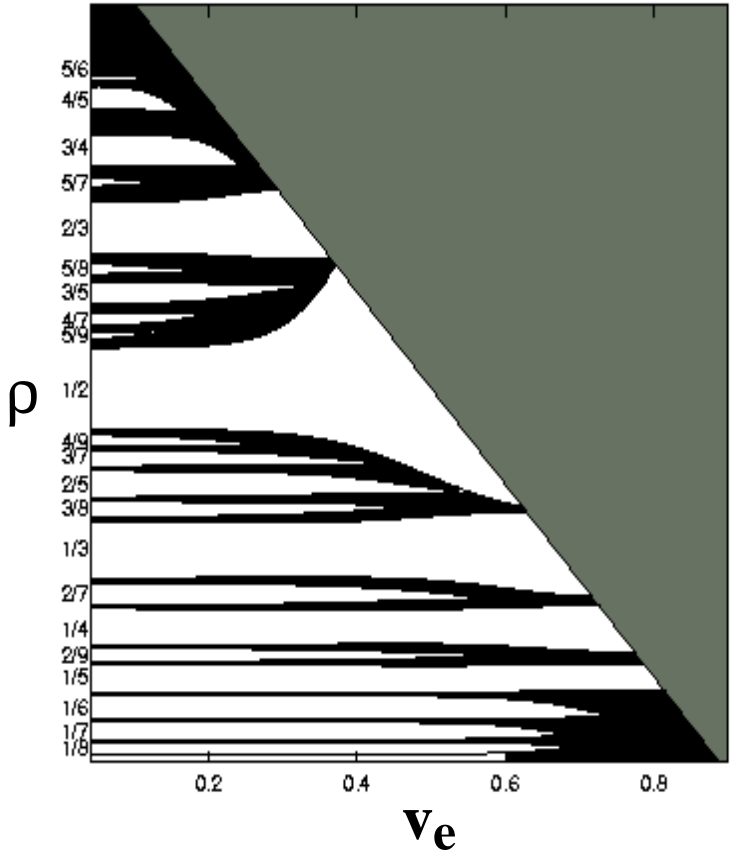}
\end{center}
\caption[]
{The dynamical phase diagram in $v_e$ and $\rho$ plane. The numbers on the 
$\rho$ axis mark the fractions $\tilde{\rho}$, which determines
the orientation of the lock-in phase. The three regions white, black and grey 
correspond to the rippled, the disordered and the detached 
phases respectively.} 
\label{pdia}
\end{figure}

As the velocity $v_e$ is increased,
steps corresponding to ${\tilde \rho}= m/n$ (rational fractions) dissappear
sequentially for $v_e > 1/m$ so that for $v_e > 1/2$ only fractions
of the form $1/n$ remain which persist upto $v_e = v_{\infty}$.
The locus of the discontinuities 
in the $\tilde{\rho}(\rho)$ curve for various velocities $v_e$ gives the 
phase diagram (Fig.~\ref{pdia}) in the $v_e - \rho$ plane. Note that the 
phase diagram for this C-I transition as given in an earlier 
publication\cite{prl} contained inaccuracies which have been now corrected. 

\subsection{ Dynamics of the forced Interface}

For low velocities and density where correlation effects due to the 
hard core constraint are negligible, the dynamics of the interface may be 
obtained exactly. Under these circumstances the $N_p$ 
particle probability distribution for the $y_i$'s, $P(y_1,y_2,\cdots,y_{N_p})$
factorizes into single particle terms $P(y_i)$. Knowing the time development
of $P(y_i)$ and the ground state structure the motion of 
the interface at subsequent times may be trivially computed as a sum of 
single particle motions. 
A single particle (with say index $i$) moves with
the bias $\Delta_i(v_e\,t)$ which, in general, may change sign at 
$y < i/\rho+v_e\,t < y + 1$. Then $P(y_i)$ satisfies the following set of
master equations,
\begin{eqnarray}
\dot P(y_i) & = & -P(y_i) + P(y_i + 1) \,\,\,\,\,\,\,\,\,\,\,\,\,\,\,\,\,\,\,\,\, {\rm for\,\,\, y_i\, >\, y+1} \nonumber\\
\dot P(y_i) & = & P(y_i - 1) - P(y_i) + P(y_i + 1)\,\,\,\,\,\,\,{\rm for\,\,\, {y_i}\, =\, y,y+1} \nonumber\\
\dot P(y_i) & = & -P(y_i) + P(y_i - 1)\,\,\,\,\,\,\,\,\,\,\,\,\,\,\,\,\,\,\,\,\, {\rm for\,\,\, {y_i}\, <\, y}.
\label{master}
\end{eqnarray}

\noindent

The average position of the particle
is given by $<y_i(t)> = \sum_{y_i=-\infty}^{\infty} y_i\,P(y_i)$ and 
the spread by $\sigma^{2}(t) = \sum_{y_i=-\infty}^{\infty} 
(y_i - <y_i(t)>)^2\,P(y_i)$.
Solving the appropriate set of master 
equations we obtain,
for $v_e << 1$ the rather obvious steady state solution 
$P(y_i) = 1/2(\delta_{y_i,y}+\delta_{y_i,y+1})$ and the particle 
oscillates between $y$ and $y+1$. Subsequently, when $i/\rho+v_e\,t \geq y+1$,
the particle jumps to the next position and $P(y_i)$ relaxes exponentially
with a time constant $\tau = 1$ to it's new value with $y \to y+1$.
For $\rho = 1/n$ the entire interface moves 
as a single particle and the average position advances in steps with a 
periodicity of $1/v_e$ (see Fig.~\ref{hops})
In general, for rational $\rho = m/n$, the motion
of the interface is composed of the independent motions of $m$ particles
each separated by a time lag of $\tau_{L} = 1/m\,v_e$.
The result of the analytic calculation for small $v_e$ and $\rho$ has been 
compared to those from simulations in Fig.~\ref{hops}  for 
$\rho = 1/5$  and $2/5 $. 
For a general irrational $\rho < 1/2$, $m \to \infty$ consequently, 
$\tau_{L} \to 0$. 

The forward motion of an irrational interface is accompanied by the motion of 
discommensurations along the interface with velocity $v_e$ which 
constitutes a kinematic wave\cite{Barabasi,Maj1} parallel to the interface. 

\subsection{The C-I transition}

\begin{figure}[t]
\begin{center}
\includegraphics[width=6.5cm]{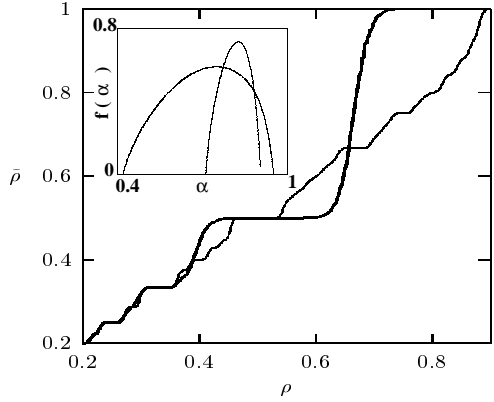}
\end{center}
\caption[]
{The function $\tilde\rho(\rho)$ for two different velocities $v_e/(1-\rho) = 
0.1$ and 
$ 0.5$. Note that the steps corresponding to higher rational fractions tend to 
disappear with increasing $v_e$. Inset shows the corresponding multifractal
spectrum $f(\alpha)$. Note that with increasing velocity, $\alpha_{min} \to 0$ 
(see text).} 
\label{falpha}
\end{figure}

As the velocity $v_e$ is increased 
the system finds it increasingly difficult to maintain
its ground state structure and for $\tau \geq \tau_L$
the instantaneous value of
$\tilde{\rho}$ begins to make excursions to other nearby low-lying fractions
and eventually becomes free. 
Steps corresponding to ${\tilde \rho}= m/n$ dissappear (i.e. 
${\tilde \rho}\to\rho$)
sequentially in order of decreasing $m$ and the interface loses the ripples.
The transition, as in the case of the FK model\cite{FK} is characterized 
by well defined exponents.
This may be seen by computing the spectrum of singularities

\begin{equation}
f(\alpha) = q \frac{d}{dq}[(q - 1)D_{q}] - (q - 1)D_{q} 
\end{equation}

\noindent
with $\alpha(q) = (d/dq)((q-1)D_{q})$ 
and $q = f^{'}(\alpha(q))$\cite{multifr} 
corresponding to the Devil's staircase in $\tilde\rho(\rho)$.
$D_{0}$ is the Hausdorff dimension and $D_{q}$ the generalized dimensions. 
$D_{q}$'s are obtained by solving for
 
\begin{equation}
\sum_{i}(p_{i}^{q}/l_{i}^{(q - 1)D_{q}}) = 1.
\end{equation}
The changes in $\rho$ determined the scales $l_{i}$ of the partition 
(defined following the Farey construction) whereas the changes in $\tilde\rho$
were defined to be the measures $p_{i}$. The high~-order gaps in the vicinity 
of primary steps (the $1/n$ fractions) scale like\cite{multifr} 
$p_i \sim l_i^{\alpha_{min}}$ where $\alpha_{min} = D_{\infty}$. Near these
steps $\tilde\rho \sim (\rho - \rho_{max})^{\xi}$ where $\rho_{max}$ is the
maximum value of $\rho$ for a step. This universal critical 
exponent\cite{modelock} $\xi = \alpha_{min}$ has been determined to be 
$0.71 \pm .001$ from our data at $v_e / (1-\rho) = 0.1$. The exponent $\xi$ 
determines the stability of the rippled pattern to small changes in the 
orientation of the external field ($\rho$). As $v_e$ increases, $\xi \to 0$
(Fig.~\ref{falpha}).

It is important to realize that the C-I transition seen
in our system is driven by fluctuations of the local slope and therefore 
different from the C-I transition in a mechanistic 
Frenkel~-Kontorova\cite{Chaikin}  model appropriate for domain structures
arising from atomic misfits. The non-local energy $E(\rho)$ and the 
non-linear constraint $y_i = {\rm integer}$ makes it extremely 
difficult to devise a natural mapping of this problem onto an effective 
F-K model. However, an approach based on the Langevin dynamics of 
particle of a single particle with coordinate $\rho^\prime$ diffusing on a
energy surface given by, $
F(\rho^\prime) = E(\rho^\prime) + \kappa\,(\rho^\prime - \rho)^2
$ can obtain the main qualitative results of this section\cite{prl}. Here, 
$\kappa$ is an arbitrary constant which ensures that the long time limit
of $\rho^\prime = \rho$.
 
\section{ CONCLUSION} 

In this paper, we have studied the static and dynamic properties of an 
Ising interface in 2-d subject to a non-uniform, time-dependent external 
magnetic field. The system has a rich dynamical structure with 
infinitely many steady states. The nature of these steady 
states and their detailed dynamics depend on the orientation of the 
interface and the velocity of the external field profile.
In future we would like to study the statics and dynamics of such forced 
interfaces in more realistic systems eg. a liquid-solid interface 
produced by coupling to a patterned substrate\cite{AMOLF}. The authors 
wish to thank S. S. Manna for discussions. A.C. acknowledges C.S.I.R. 
Govt. of India for a fellowship and S.S. thanks the Department of Science 
and Technology, Govt. of India, for financial support.

 

\end{document}